\documentclass[sn-mathphys-num]{sn-jnl}
\usepackage{graphicx}%
\usepackage{multirow}%
\usepackage{amsmath,amssymb,amsfonts}%
\usepackage{amsthm}%
\usepackage{mathrsfs}%
\usepackage[title]{appendix}%
\usepackage{xcolor}%
\usepackage{textcomp}%
\usepackage{manyfoot}%
\usepackage{booktabs}%
\usepackage{algorithm}%
\usepackage{algorithmicx}%
\usepackage{algpseudocode}%
\usepackage{listings}%
\usepackage{esint}%
\usepackage{mathtools}%

\begin{document}

\title[Article Title]{Physics of radio antennas}

\author*[1]{\fnm{Mohammad Ful Hossain} \sur{Seikh}}\email{fulhossain@ku.edu}

\affil*[1]{\orgdiv{Department of Physics \& Astronomy}, \orgname{University of Kansas}, \orgaddress{\city{Lawrence}, \postcode{66045}, \state{KS}, \country{USA}}}


\abstract{
Radio antennas are widely used in the field of particle astrophysics in searches for ultra-high energy cosmic rays (UHECR) and neutrinos (UHEN). It is therefore necessary to properly describe the physics of their response. In this article, we summarize the mathematics underlying parameterizations of radio antennas. As a paradigm, we focus on a half-wave dipole and also discuss measurements of characteristics, performed in an electromagnetic (EM) anechoic chamber.
}

\maketitle

\section{Introduction}\label{sec1}

Radio neutrino and cosmic ray experiments such as the Askaryan Radio Array (ARA) \cite{1}, Radio Neutrino Observatory-Greenland (RNO-G) \cite{2}, Radar Echo Telescope (RET) \cite{3}, Pierre Auger Observatory (PAO) \cite{7} and others deploy radio antennas as the front-end sensor for their detectors. Those antennas are often sensitive over a wide bandwidth, starting from 50 MHz and extending to a few GHz, making them ideal for UHEN and UHECR observation through Askaryan radiation, geomagnetic radiation, and radar echoes. These techniques are based on the detection of coherently superposed  photons, each with meter-scale wavelengths  (frequency $\sim \mathcal{O}(100)$ MHz)  rather than detection of incoherently summed photons or high frequency optical single photons. Additionally, radio antennas are robust and relatively simple and cheap to fabricate. Radio antennas can be used either as a transmitter (Tx), in the case when EM signal arising from electrical current in its conductor is radiated, or a receiver (Rx) when it receives EM radiation and converts it into electrical current.\\\\
Before discussing the primary antenna parameters, we review some basic electromagnetic theory. We start with the free-space Maxwell’s equations in a source-free region (e.g., a vacuum). The fields (function of position, $\Vec{r}$ and time, $t$) in a region far away from the current or charges (sources) that created them can be written in differential equation form as 
\begin{equation}
\Vec{\nabla} \cdot \Vec{E} = 0,~~ \Vec{\nabla} \times \Vec{E} = -\frac{\partial \Vec{B}}{\partial t}, ~~\Vec{\nabla} \cdot \Vec{B} = 0, ~~ \Vec{\nabla} \times \Vec{B} = \mu_0\varepsilon_0\frac{\partial \Vec{E}}{\partial t}. \label{eqn1}
\end{equation}
Following the standard derivation, and taking the curl of Faraday's law (second Eqn. of \eqref{eqn1}), we obtain the vector wave equation for the electric field (Volts/m) as
\begin{align}
        \Vec{\nabla} \times (\Vec{\nabla} \times \Vec{E}) = -\frac{\partial (\Vec{\nabla} \times \Vec{B})}{\partial t} ~~~~& \Longrightarrow~~~~
        \Vec{\nabla} (\Vec{\nabla} \cdot \Vec{E}) - \nabla ^2 \Vec{E} = - \mu_0\varepsilon_0\frac{\partial^2 \Vec{E}}{\partial t^2} \nonumber \\
        & \Longrightarrow~~~~ \nabla ^2 \Vec{E} - \mu_0\varepsilon_0\frac{\partial^2 \Vec{E}}{\partial t^2} = 0. \label{eqn2}
\end{align}
The electric field cannot be any arbitrary function of $\Vec{r}$ and $t$. Instead, it is physically possible only if it satisfies the differential Eqn. \eqref{eqn2}. The simplest solution is the plane-wave solution where the electric field varies with time in a sinusoidal manner with an angular frequency of $\omega$ radians/sec. Note that this field is a function of spatial coordinate, $z$, but the direction of the electric field is orthogonal to the z-axis.
\begin{align}
\Vec{E} = ({E_x}\hat{x} + {E_y}\hat{y})e^{j\phi} = ({E_x}\hat{x} + {E_y}\hat{y})e^{j\omega(t - z\sqrt{\mu_0\varepsilon_0})} 
\label{eqn3}
\end{align}
How fast does the wave move? For a constant phase $\phi$, 
\begin{align}
\frac{d\phi}{dt} = 0 ~~~~\Longrightarrow~~~~ \frac{dz}{dt} = v_p = \frac{1}{\sqrt{\mu_0\varepsilon_0}} = c. 
\label{eqn4}
\end{align}
Therefore, E-field moves with a phase velocity, $v_p$, equal to the speed of light, c. Similarly, the magnetic field also moves with c. From Eqns. \eqref{eqn3} and \eqref{eqn4},
\begin{equation}
    \Vec{E} = ({E_x}\hat{x} + {E_y}\hat{y})e^{j\omega t} e^{ -j (\omega/c)z} \label{eqn5}, 
\end{equation}
where $(\omega/c) = k_0 = 2\pi/\lambda$ is called the `spatial frequency'. The wave vector along the propagation direction is $\Vec{k} = k_0 \hat{z}$. There are many other solutions besides the plane-wave solution. The most relevant of these, perhaps, is the spherical wave,
\begin{equation}
    \Vec{E} = ({E_\theta}\hat{\theta} + {E_\phi}\hat{\phi}) \frac{e^{ - jk_0 r}}{r} e^{j\omega t}. \label{eqn6}
\end{equation} 
Similar to the electric field wave Eqn. \eqref{eqn2}, the magnetic field (Webers/m$^2$) wave equation can be written as 
\begin{equation}
   \nabla ^2 \Vec{B} - \mu_0\varepsilon_0\frac{\partial^2 \Vec{B}}{\partial t^2} = 0. \label{eqn7}
\end{equation}
The $\Vec{E}$, $\Vec{B}$, and $\Vec{k}$ are mutually perpendicular and comprise an orthogonal set of vectors; hence, the `transverse' nature of EM waves. From the known E-field, we can write the B-field as
\begin{equation}
\Vec{B} = ({B_x}\hat{x} + {B_y}\hat{y})e^{j\omega t} e^{ - jk_0 z} = \frac{k_0}{\omega}(- {E_y}\hat{x} + {E_x}\hat{y})e^{j\omega t} e^{ - jk_0 z}. \label{eqn8} 
\end{equation}
In addition, we define the magnetizing field (Amperes/m) as
\begin{equation}
\displaystyle{\Vec{H} = \frac{\Vec{B}}{\mu_0} = \frac{k_0}{\mu_0\omega}(- {E_y}\hat{x} + {E_x}\hat{y})e^{j\omega t} e^{ - jk_0 z}}.
\label{eqn9} 
\end{equation}
Now we can derive two important ratios.
\begin{equation}
\frac{|\Vec{E}|}{|\Vec{B}|} = \frac{\omega}{k_0} \frac{\sqrt{|E_x|^2 + |E_y|^2}}{\sqrt{|E_y|^2 + |E_x|^2}} = \frac{k_0c}{k_0} = c,
\label{eqn10} 
\end{equation}
and 
\begin{equation}  
\frac{|\Vec{E}|}{|\Vec{H}|} = \frac{\mu_0\omega}{k_0} \frac{\sqrt{|E_x|^2 + |E_y|^2}}{\sqrt{|E_y|^2 + |E_x|^2}} = \mu_0c = \frac{\mu_0}{\sqrt{\mu_0\varepsilon_0}} = \sqrt{\frac{\mu_0}{\varepsilon_0}} = \eta_0.
\label{eqn11} 
\end{equation}
The first ratio motivates why the radio neutrino and cosmic ray experiments measure the electric, and not magnetic field. The second ratio is called the free space wave impedance, $Z_f = \eta_0$ which is equal to $120\pi~\Omega = 377~\Omega$.

\section{Antenna Parameters}\label{sec2}
The power radiated per unit area is a real-valued vector called the Poynting vector, corresponding to a radiation energy density (Watts/m$^2$). It is defined as
\begin{align}
    \Vec{W} = \frac{1}{2} \Re{\left(\Vec{E} \times \Vec{H^*}\right)}
    & = \frac{1}{2 Z_f} \Re{\left(({E_x}\hat{x} + {E_y}\hat{y}) \times (- {E_y^*}\hat{x} + {E_x^*}\hat{y})\right)}\nonumber\\
    & = \frac{1}{2 Z_f} {\left(|E_x|^2 + |E_y|^2\right)}\hat{z} = \frac{\hat{z}}{2 Z_f}|\Vec{E}|^2. \label{eqn12} 
\end{align}
    Generally, in a medium with refractive index n,
\begin{align}
|\Vec{W}| = \frac{|\Vec{E}|^2}{2Z_f/n} = \frac{|\Vec{E}|^2}{240\pi}n. \label{eqn13} 
\end{align}

\subsection{Radiation Intensity}\label{subsec2.1}
A radiating antenna launches spherical waves. In spherical coordinates, the radiation density can be written as
\begin{align}
\Vec{W} = \frac{\hat{r}}{r^2} \frac{|E_\theta (\hat{r})|^2 + |E_\phi (\hat{r})|^2}{2 Z_f}. \label{eqn14}
\end{align}
We define a real-valued scalar function 
\begin{align}
U(\hat{r}) =  \frac{|E_\theta (\hat{r})|^2 + |E_\phi (\hat{r})|^2}{{2 Z_f}} = \frac{|\Vec{E} (\hat{r})|^2}{2Z_f}, \label{eqn15}
\end{align}
as the radiation intensity (Watts/sr) of a spherical wave. $U(\hat{r}) = U(\theta, \phi)$ describes the areal distribution of power, i.e., it prescribes how the radiated power is distributed as a function of radiation direction. Therefore, the radiation density can be written as 
\begin{align}
\Vec{W} =  U(\hat{r}) \frac{\hat{r}}{r^2}. \label{eqn16}
\end{align}
This Eqn. \eqref{eqn16} helps  us understand various antenna characterizing parameters. 

\subsection{Radiated Power}\label{subsec2.2}
Through any closed surface, the power radiated by an antenna is
\begin{align}
P_{rad} &= \varoiint_{s} \Vec{W}\cdot\Vec{dS} = \varoiint_{s} U(\theta, \phi)\frac{\hat{r}}{r^2} \cdot \hat{r}r^2\sin\theta d\theta d\phi \nonumber\\
& = \int_\theta \int_\phi U(\theta, \phi)\sin\theta d\theta d\phi = \int_\theta \int_\phi U(\theta, \phi)d\Omega. \label{eqn17}
\end{align}
Can an antenna radiate equal power in all direction? Yes, a `mythical' monopole isotropic radiator can, with a constant radiation intensity
\begin{align}
U_0 = U(\theta, \phi)  = \frac{P_{rad}}{\int_{\theta = 0}^{\pi} \int_{\phi = 0}^{2\pi} d\Omega} = \frac{P_{rad}}{4\pi}. \label{eqn18}
\end{align}

\subsection{Directivity Pattern}\label{subsec2.3}
Unlike an isotropic radiator, a real antenna does not radiate power equally in all directions. The pattern of power distribution of an antenna over the full solid angle with respect to the same power radiated by an isotropic radiator is called the antenna directivity pattern, 
\begin{align}
D(\hat{r}) = \frac{\mbox{intensity~of~antenna}}{\mbox{intensity~ of~isotropic~radiator}} = \frac{U(\hat{r})}{U_0} = \frac{4\pi U(\hat{r})}{P_{rad}}. \label{eqn19}
\end{align}
The unitarity condiction on the function given in Eqn. \eqref{eqn19} is 
\begin{align}
\int\int D(\hat{r}) d\Omega = \frac{4\pi}{P_{rad}}\int\int U(\hat{r})d\Omega = 4\pi. \label{eqn20}
\end{align}
Therefore, the average directivity over the full solid angle of any and all antennas must be equal to 1. The directivity pattern is often symmetric over some zenith ($\theta$) and/or azimuth ($\phi$) angles as antenna geometry, in general, are symmetric over some plane. The region in the directivity pattern with maximum directivity value is called the antenna mainlobe. 

\subsubsection{ Directivity}\label{subsubsec2.3.1}
Antenna `directivity' is defined as the maximum value in the directivity pattern
\begin{align}
D_0 = max\{D(\theta, \phi)\}. \label{eqn21}
\end{align}
In practice, it is expressed in dB units as
\begin{align}
dB[D_0] = 10\log_{10}[D_0]. \label{eqn22}
\end{align}
\subsubsection{ Beamwidth}\label{subsubsec2.3.2}   
`Beamwidth' is defined by the region of the mainlobe solid angle for which the directivity pattern has a value of at least $1/2$ that of $D_0$ (i.e., no more than 3 dB smaller than $dB[D_0]$). It is denoted by $\Omega_A$ and is a product of the beamwidth in the zenith ($\beta_{\theta}$) and azimuthal ($\beta_\phi$) planes. If the mainlobe contains most of the radiated power, we can approximate 
\begin{align}
\Omega_A = \beta_{\theta} \beta_{\phi} = \iint\limits_\mathrm{mainlobe} d\Omega \approx \frac{4\pi}{D_0}. \label{eqn23}
\end{align}
\subsubsection{Peak~Sidelobe~Level}\label{subsubsec2.3.3} 
In addition to the mainlobe, there can be several sidelobes in the directivity pattern. The peak sidelobe level is defined as   
\begin{align}
dB[PSL] = 10\log_{10}[D_{sl}/D_0] ; ~~~~ D_{sl} = max\{D(\theta, \phi)_{side~lobes}\}. \label{eqn24}
\end{align}

\subsection{Antenna Impedance}\label{subsec2.4}
The main difference between an antenna and a resistor is that a resistor converts its absorbed power into  heat whereas an antenna (ideally) converts its absorbed power into a propagating, spherical,  EM wave. When a power cable (carrying $P_{avl}^{Tx}$ power from Tx) having characteristic impedance $Z_0$ is connected to an antenna with impedance $Z_A$, there is an impedance mismatch which leads to reflection of some EM energy at the junction. This reflection coefficient is given by
\begin{align}
    \Gamma_A = \frac{Z_A - Z_0}{Z_A + Z_0}. \label{eqn25}
\end{align}
This complex antenna impedance, like all other antenna parameters, is frequency dependent
\begin{align}
Z_A (f) = R_A (f) + jX_A (f). \label{eqn26}
\end{align}
Looking at the imaginary part of the impedance, we can divide frequency domains for which the response is primarily inductive ($Z_A$ positive; $X_A\sim\omega L_A$) or capacitive ($Z_A$ negative; $X_A\sim-1/\omega C_A$). Here, $R_A$ is the resistive part of the complex antenna impedance.

\subsubsection{ Efficiency}\label{subsubsec2.4.1} 
Most of the power available from the Tx is absorbed by an antenna, but some of the available power is reflected at the cable-antenna junction. Most of the absorbed power ($P^A_{abs}$, $R_A$) is radiated ($P_{rad}$, $R_{rad}$), but some of the absorbed power is converted to heat and is lost ($P_L$, $R_{L}$) via Joule heating. Thus, various powers and their associated resistances can be expressed as
\begin{align}
P^A_{abs} = P_{avl}^{Tx}\left(1-|\Gamma_A|^2\right) = P_L + P_{rad} ~~~~\Longrightarrow~~~~ R_A = R_L + R_{rad}. \label{eqn27}
\end{align}   
Such losses within an antenna determine its efficiency, $e\le 1$, defined as
\begin{align}
e = \frac{P_{rad}}{P^A_{abs}} = \frac{P^A_{abs} - P_L}{P^A_{abs}} = 1 - \frac{P_L}{P^A_{abs}} = 1 - \frac{R_L}{R_A}. \label{eqn28}
\end{align}
A part of $P_L$ is dissipated as dielectric loss {($P_{dl}$)}, a part as conductive loss {($P_{cl}$)}, and some fraction may be dissipated as unwanted surface waves {($P_{sw}$)} in the antenna, with corresponding efficiencies associated with these individual loss mechanisms as $e_d$, $e_c$, and $e_{sw}$, respectively. Correspondingly,
\begin{align}
P_L = P_{dl} + P_{cl} + P_{sw}. \label{eqn29}
\end{align}
\begin{equation}
        \therefore~~e = 1 - \frac{P_{dl} + P_{cl} + P_{sw}}{P^A_{abs}} \simeq \left( 1 - \frac{P_{dl}}{P^A_{abs}} \right) \left( 1 - \frac{P_{cl}}{P^A_{abs}} \right) \left( 1 - \frac{P_{sw}}{P^A_{abs}} \right) = e_de_c e_{sw}. \label{eqn30}
\end{equation}

\subsection{Gain Pattern}\label{subsec2.5}
The gain of a transmitting antenna in a given direction is the ratio of the broadcast radiation intensity in that direction to the power absorbed by the antenna from the feedpoint (absorbed from source),
\begin{equation}
        G(\theta, \phi) = 4\pi \frac{U(\theta, \phi)}{P^A_{abs}} ~~~~\Longrightarrow~~~~G = eD. \label{eqn31}
\end{equation} 

\subsection{Realized Gain Pattern}\label{subsec2.6}
The realized gain of a transmitting antenna in a given direction is the ratio of the broadcast radiation intensity in that direction to the input power at the antenna feedpoint (available from source),
\begin{equation}
        G^r(\theta, \phi) = 4\pi \frac{U(\theta, \phi)}{P_{avl}^{Tx}} ~~~~\Longrightarrow~~~~G^r = \left(1-|\Gamma_A|^2\right)G = e\left(1-|\Gamma_A|^2\right)D. \label{eqn32} 
\end{equation} 

\subsection{Antenna Polarization}\label{subsec2.7}
In typical antenna applications, a transmitting antenna broadcasts signal to a receiving antenna. Let us assume a Tx antenna is located at $\Vec{r}^\prime$, and we would like to measure the electric field at an arbitrary position $\Vec{r}$. The direction of propagation in this case will be denoted as the unit vector
\begin{equation}
\hat{k} = \frac{\Vec{r}  - \Vec{r}^\prime}{|\Vec{r}  - \Vec{r}^\prime|}. \label{eqn33} 
\end{equation} 
If the transmitter is located at the origin (i.e., $\Vec{r}^\prime = 0$), the unit vector $\hat{k} = \hat{r}$ and we have polarization ($\hat{p}$) of electric field along $\hat{\theta}$ and/or $\hat{\phi}$. In the more general case of $\Vec{r}^\prime \neq 0$, $\hat{\theta}\cdot\hat{k} \neq \hat{\phi}\cdot\hat{k} \neq 0$. We can construct three mutually orthonormal vectors; $\hat{k}, \hat{a_2}, \hat{a_3}$ such that
$\hat{a_2}\cdot\hat{k} = \hat{a_3}\cdot\hat{k} = \hat{a_2}\cdot \hat{a_3} = 0$. We follow the standard procedure for picking $\hat{a_2}$ and $\hat{a_3}$, taking advantage of the fact that two spherical basis vectors, $\hat{r}$ and $\hat{\theta}$ have components in the direction of all three Cartesian basis vectors but not $\hat{\phi}$, 
\begin{align}
\hat{r} &= \sin\theta\cos\phi \hat{x} + \sin\theta\sin\phi \hat{y} + \cos\theta \hat{z},\nonumber \\
\hat{\theta} &= \cos\theta\cos\phi \hat{x} + \cos\theta\sin\phi \hat{y} - \sin\theta \hat{z},\nonumber \\
\hat{\phi} & = -\sin\phi \hat{x} + \cos\phi \hat{y}. \label{eqn34}
\end{align}    
Thus, the basis vector $\hat{\phi}$ distinguishes itself as being orthogonal to the z-axis (generally taken to be vertically up; zenith). The horizontal polarization vector can now be written as $\hat{a_2} = \hat{h}$ which satisfies $\hat{h}\cdot \hat{k} = \hat{h} \cdot \hat{z} = 0$
\begin{equation}
\therefore~~~\hat{h} = \frac{\hat{z}\times \hat{k}}{|\hat{z}\times \hat{k}|}. \label{eqn35}
\end{equation}
Note that $\hat{h}$ is normalized with $|\hat{z}\times \hat{k}|$ as $\hat{z}$ and $\hat{k}$ are not necessarily orthogonal. In contrast to the horizontal base vector, the unit vector $\hat{a_3}$ is described as the vertical polarization unit vector
\begin{equation}
\hat{v} = \hat{k} \times \hat{h}. \label{eqn36}
\end{equation}
Thus, the unit vectors $\hat{k}, \hat{h}, \hat{v}$ form a (right-handed) orthonormal set of basis vectors; one that is independent of the location of the origin. As expected, when the Tx is located at the origin ($\Vec{r}^\prime = 0$), these vectors reduce to 
\begin{equation}
\hat{k} = \hat{r}, ~~~~ \hat{h} = \hat{\phi},~~~~ \hat{v} = -\hat{\theta}.
\label{eqn37}
\end{equation}

\subsection{Effective Aperture}\label{subsec2.8}
The receiving antenna acts as a device for coupling a propagating EM wave into the receiver input. The available power at the receiver ($P^{Rx}_{avl}$) is proportional to the magnitude of the incident radiation density
\begin{equation}
P^{Rx}_{avl} \propto |\Vec{W}_{inc}|  ~~~~\Longrightarrow~~~~ P^{Rx}_{avl} = A_e |\Vec{W}_{inc}|. \label{eqn38}
\end{equation}
$A_e$, known as the `effective aperture' of an antenna or simply the `antenna aperture', depends on a set of factors including the antenna design, the direction of the propagating wave, the frequency of the EM wave, and the wave polarization. $A_e$ is closely related to the gain pattern of the receiving antenna. In fact, we can correctly conclude that the antenna gain pattern, and effective aperture pattern are directly proportional:
\begin{align}
A_e(\hat{k}_{inc}, \hat{p}_{inc}) & \propto G(\hat{k} = -\hat{k}_{inc})~~~~\Longrightarrow~~~~ A_e = \frac{\pi}{k_0^2}|\hat{p}(\hat{k} = -\hat{k}_{inc})\cdot \hat{p}_{inc}|^2 G\nonumber\\
\therefore ~~A_e & = \frac{\lambda^2}{4\pi}|\hat{p}(\hat{k} = -\hat{k}_{inc})\cdot \hat{p}_{inc}|^2 G.
\label{eqn39}
\end{align}
\subsubsection{ Polarization Loss Factor}\label{subsubsec2.8.1}
The term $|\hat{p}(\hat{k} = -\hat{k}_{inc})\cdot \hat{p}_{inc}|^2$ is known as the polarization loss factor. It arises due to the polarization mismatch of an antenna and the received EM wave. Since the polarization vectors are unit vectors, we can conclude
\begin{equation}
0\le PLF = |\hat{p}(\hat{k} = -\hat{k}_{inc})\cdot \hat{p}_{inc}|^2 \le 1. \label{eqn40}
\end{equation}
$\therefore$ $PLF = 1$ for co-polarization and $PLF = 0$ for cross polarization.

\section{The Friis Transmission Equation}\label{sec3}
In the field of antenna physics, the Friis transmission equation (ultimately rooted in energy conservation) is one of the most important concepts. It basically establishes a relationship between the power available at a transmitting antenna, $P^{Tx}_{avl}$ and how much of it is absorbed by a receiving antenna, $P^{Rx}_{abs}$, separated by a distance $R$. We use the pedagogy of section \ref{sec2} to derive it. We need: \\\\
Power radiated from the Tx antenna (from Eqns. \eqref{eqn27} and \eqref{eqn28}),
\begin{equation}
P^{Tx}_{rad} = e\left(1-|\Gamma^{Tx}|^2\right)P^{Tx}_{avl}. \label{eqn41}
\end{equation}
Radiation intensity at the Tx antenna (from  Eqn. \eqref{eqn19}),
\begin{equation}
U^{Tx} = D^{Tx} U_0 = \frac{D^{Tx} e\left(1-|\Gamma^{Tx}|^2\right)P^{Tx}_{avl}}{4\pi}.
\label{eqn42}
\end{equation}
Radiation density at the Rx antenna (from Eqn. \eqref{eqn16}),
\begin{equation}
\Vec{W}^{Rx} =  U^{Tx} \frac{\hat{k}}{R^2} = \hat{k} \frac{D^{Tx} e(1-|\Gamma^{Tx}|^2)}{4\pi R^2} P^{Tx}_{avl}.
\label{eqn43}
\end{equation}
Available power at the Rx antenna (from Eqn. \eqref{eqn38}), 
\begin{align}
P^{Rx}_{avl} = A^{Rx}_e|\Vec{W}^{Rx}|.
\label{eqn44}
\end{align}
Absorbed power at the Rx antenna (from Eqn. \eqref{eqn27}), 
\begin{align}
P^{Rx}_{abs}  &= (1-|\Gamma^{Rx}|^2)P^{Rx}_{avl}\nonumber\\ 
&= (1-|\Gamma^{Rx}|^2)A^{Rx}_e|\Vec{W}^{Rx}|\nonumber\\
& = (1-|\Gamma^{Rx}|^2)A^{Rx}_e \frac{D^{Tx} e(1-|\Gamma^{Tx}|^2)}{4\pi R^2} P^{Tx}_{avl}\nonumber\\
\therefore~~\frac{P^{Rx}_{abs}}{P^{Tx}_{avl}} &= (1-|\Gamma^{Rx}|^2) (1-|\Gamma^{Tx}|^2)  \frac{A^{Rx}_e eD^{Tx}}{4\pi R^2}.
\label{eqn45}
\end{align}
In terms of realized gains (from Eqns. \eqref{eqn32} and \eqref{eqn39}),
\begin{equation}
    G^r_{Tx} = e(1-|\Gamma^{Tx}|^2)D^{Tx} = (1-|\Gamma^{Tx}|^2) G_{Tx}, \label{eqn46}
\end{equation} 
and 
\begin{equation}
A^{Rx}_e = \frac{\lambda^2}{4\pi}(PLF)G_{Rx} = \frac{\lambda^2}{4\pi}(PLF)\frac{G^r_{Rx}}{(1-|\Gamma^{Rx}|^2)}. \label{eqn47}
\end{equation}
Using Eqns. \eqref{eqn46} and \eqref{eqn47} in \eqref{eqn45},
\begin{align}
    \frac{P^{Rx}_{abs}}{P^{Tx}_{avl}} = \left(\frac{\lambda}{4\pi R}\right)^2(PLF) G^r_{Tx}G^r_{Rx}.\label{eqn48}
\end{align}
Eqn. \eqref{eqn48} is the most popular form of the Friis transmission equation. 

\section{Half-wave Dipole Antenna: An Example}\label{sec4}
An antenna whose total length ($2l$) is equal to half the wavelength ($\lambda/2$) of EM wave that it transmits or receives is called a `half-wave' antenna. Most radio neutrino and cosmic ray experiments use dipole antennas for the detection of UHEN and UHECR. In this section, we briefly discuss the properties of a dipole antenna. For a half-wave dipole antenna,
\begin{equation}
    k_0l = \frac{2\pi}{\lambda}l = \frac{\pi}{2}, ~~\Longrightarrow~~ l = \frac{\lambda}{4}~~\Longrightarrow~~ 2l = \frac{\lambda}{2}. \label{eqn49}
\end{equation}
The vector potential field around a half wave dipole \cite{4}\cite{5}, oriented along the z-axis is given by
\begin{equation}
\left.\Vec{A}(\Vec{r})\right|_{k_0l = \pi/2} = \hat{z}\frac{I_0}{2\pi}\frac{e^{-jk_0 r}}{k_0 r} \frac{\cos(\frac{\pi}{2}\cos\theta)}{\sin^2\theta}. \label{eqn50}
\end{equation}
Using Eqn. \eqref{eqn50}, the electric field can be derived as
\begin{align}
     \Vec{E}(\Vec{r}) & = \frac{1}{j\omega\varepsilon_0}\Vec{\nabla}\left(\Vec{\nabla} \cdot \Vec{A}(\Vec{r})\right) - j\omega\mu_0 \Vec{A}(\Vec{r}) \nonumber\\
     &= \hat{r} \frac{I_0}{2\pi} e^{-jk_0 r}\left(j\frac{2Z_f}{k_0^2}\textcolor{blue}{\frac{1}{r^3}}\right)\frac{\cos\theta\cos(\frac{\pi}{2}\cos\theta)}{\sin^2\theta}\nonumber\\
      &+ \hat{r} \frac{I_0}{2\pi} e^{-jk_0 r}\left( \frac{Z_f}{k_0}\textcolor{red}{\frac{1}{r^2}} - j\frac{2Z_f}{k_0^2}\textcolor{blue}{\frac{1}{r^3}}\right)\frac{\pi}{2}\sin\left(\frac{\pi}{2}\cos\theta\right)\nonumber\\
      &+ \hat{\theta} \frac{I_0}{2\pi} e^{-jk_0 r}\left(\frac{Z_f}{k_0}\textcolor{red}{\frac{1}{r^2}}-j\frac{Z_f}{k_0^2}\textcolor{blue}{\frac{1}{r^3}}\right)\left[\frac{\pi}{2}\cot\theta\sin\left(\frac{\pi}{2}\cos\theta\right) -(1+2\cot^2\theta)\frac{\cos(\frac{\pi}{2}\cos\theta)}{\sin\theta}\right]\nonumber\\
      &+ \hat{\theta} \frac{I_0}{2\pi} e^{-jk_0 r}\left(-j\frac{Z_f}{k_0^2}\textcolor{blue}{\frac{1}{r^3}}\right) \left[\left(\frac{\pi}{2}\right)^2\cos\left(\frac{\pi}{2}\cos\theta\right)\sin\theta\right]\nonumber\\
      &+\hat{\theta}\frac{I_0}{2\pi}e^{-jk_0 r}\left(j Z_f \textcolor{green!75!blue}{\frac{1}{r}}\right) \frac{\cos(\frac{\pi}{2}\cos\theta)}{\sin\theta}. \label{eqn51}
\end{align}
Similarly, the magnetizing field can be calculated as
\begin{align}
     \Vec{H}(\Vec{r}) &= \Vec{\nabla} \times \Vec{A}(\Vec{r}) \nonumber\\
     &=\hat{\phi} \frac{I_0}{2\pi}\frac{\cos(\frac{\pi}{2}\cos\theta)}{\sin^2\theta} e^{-jk_0 r}\left(\frac{1}{k_0}\textcolor{red}{\frac{1}{r^2}} + j\textcolor{green!75!blue}{\frac{1}{r}}\right)\sin\theta\nonumber\\
      &- \hat{\phi} \frac{I_0}{2\pi}e^{-jk_0 r} \left(\frac{1}{k_0}\textcolor{red}{\frac{1}{r^2}}\right)\left[\frac{\frac{\pi}{2}\sin(\frac{\pi}{2}\cos\theta)}{\sin\theta}-2\left(\frac{\cos\theta\cos(\frac{\pi}{2}\cos\theta)}{\sin^3\theta}\right)\right]\cos\theta. \label{eqn52}
\end{align}
Note that the magnetizing field is directed towards $\hat{\phi}$ which follows the right-hand rule; current flows through a dipole along z-axis. In experiments with radio antennas, we typically consider receiving and transmitting antennas (maximum dimension, $D$ $\sim$ $\mathcal{O}$(1) m) separated by a distance $R \sim \mathcal{O}$(10) m far from each other. The electric and magnetizing fields around them are not isotropically distributed. Some field components decay rapidly with distance (confined to a region close to the antenna, with often a circular geometry) whereas others propagate over long distances. Correspondingly, fields (at a distance $r$) are divided into three regions (discussed elaborately in \cite{5}):\\\\
Region 1: reactive near field;  $\displaystyle{0<r<0.62\sqrt{D^3/\lambda}}$\\
Region 2: radiative near field (Fresnel region);  $\displaystyle{0.62\sqrt{D^3/\lambda}\le r < 2D^2/\lambda}$\\
Region 3: radiative  far field (Fraunhoffer region);  $\displaystyle{2D^2/\lambda\le r \le \infty}$\\\\
In the field Eqns. \eqref{eqn51} and \eqref{eqn52}, only one term each (associated with $\textcolor{green!75!blue}{1/r}$) contribute to the Poynting vector, corresponding to the field propagating away from the antenna. These terms are called  `far field' terms. Terms associated with $\textcolor{red}{1/r^2}$ and $\textcolor{blue}{1/r^3}$  describe the field components that swirl around and diminish quickly as we move away from the antenna. These terms are called  `near field' terms. Therefore, the fundamental far fields (from Eqns. \eqref{eqn51} and \eqref{eqn52}), are
\begin{align}
     \Vec{E}_{far}(\Vec{r}) &= \hat{\theta}\frac{I_0}{2\pi}e^{-jk_0 r}\left(j Z_f \textcolor{green!75!blue}{\frac{1}{r}}\right) \frac{\cos(\frac{\pi}{2}\cos\theta)}{\sin\theta}, \label{eqn53}\\
     \Vec{H}_{far}(\Vec{r}) &= \hat{\phi}\frac{I_0}{2\pi}e^{-jk_0 r}\left(j \textcolor{green!75!blue}{\frac{1}{r}}\right) \frac{\cos(\frac{\pi}{2}\cos\theta)}{\sin\theta}. \label{eqn54}
\end{align}
Comparing with the spherical wave fields
\begin{align}
\Vec{E}(\Vec{r}) & = ({E_\theta}\hat{\theta} + {E_\phi}\hat{\phi}) \frac{e^{ - jk_0 r}}{r}, \label{eqn55}\\
\Vec{H}(\Vec{r}) & = ({H_\theta}\hat{\theta} + {H_\phi}\hat{\phi}) \frac{e^{ - jk_0 r}}{r}, \label{eqn56}
\end{align}
we can infer
\begin{equation}
E_\phi = 0 = H_\theta, \label{eqn57}
\end{equation}
and 
\begin{equation}
E_\theta = jZ_f\frac{I_0}{2\pi} \frac{\cos(\frac{\pi}{2}\cos\theta)}{\sin\theta} = Z_f H_\phi. \label{eqn58}
\end{equation}

\subsection{Antenna Polarization}\label{subsec4.1}
The polarization unit vector (using Eqns. \eqref{eqn57} and \eqref{eqn58}), is given by
\begin{equation}
    \hat{p}(\hat{r}) = \frac{{E_\theta}\hat{\theta} + {E_\phi}\hat{\phi}}{\sqrt{|{E_\theta}|^2 + |{E_\phi}|^2}} = j\hat{\theta}. \label{eqn59}
\end{equation}
Using Eqn. \eqref{eqn37} in \eqref{eqn59}, we obtain the dipole polarization as
\begin{equation}
    \hat{p}(\hat{r}) = j\hat{\theta} = -j\hat{v}. \label{eqn60}
\end{equation}
Therefore, a vertical dipole is vertically (linearly) polarized. 

\subsection{Radiated Power}\label{subsec4.2}
The Poynting vector (or, radiation density) for a half-wave dipole (from Eqns. \eqref{eqn53}, \eqref{eqn54} and \eqref{eqn58}) is
\begin{align}
    \Vec{W} &= \frac{1}{2} \Re{\left(\Vec{E}_{far} \times \Vec{H^*}_{far}\right)} = \frac{(\hat{\theta}\times\Vec{\phi})}{r^2}\frac{1}{2Z_f}|\Vec{E}|^2 = \frac{\hat{r}}{r^2}\frac{1}{2Z_f}(|{E_\theta}|^2+{E_\phi}|^2)\nonumber\\
    &=\frac{\hat{r}}{r^2}\frac{1}{2Z_f}\left|jZ_f\frac{I_0}{2\pi} \frac{\cos(\frac{\pi}{2}\cos\theta)}{\sin\theta}\right|^2 = \frac{|I_0|^2}{8\pi^2}Z_f \left(\frac{\cos(\frac{\pi}{2}\cos\theta)}{\sin\theta}\right)^2 \frac{\hat{r}}{r^2} = U(\hat{r})\frac{\hat{r}}{r^2}\nonumber\\
    \therefore ~~U(\hat{r}) &= \frac{|I_0|^2}{8\pi^2}Z_f \left(\frac{\cos(\frac{\pi}{2}\cos\theta)}{\sin\theta}\right)^2.
    \label{eqn61}
\end{align}
Therefore, the radiated power is given by 
\begin{align}
       P_{rad} &= \int_0^{2\pi}\int_0^\pi U(\hat{r}) \sin\theta d\theta d\phi\nonumber\\
       &= \frac{|I_0|^2}{8\pi^2}Z_f  \int_0^{2\pi} d\phi \int_0^\pi  \frac{\cos^2(\frac{\pi}{2}\cos\theta)}{\sin\theta} d\theta\nonumber\\
       &\simeq\frac{|I_0|^2}{4\pi}Z_f(1.22). \label{eqn62}
\end{align}

\subsection{Directivity Pattern}\label{subsec4.3}
Using Eqns. \eqref{eqn61} and \eqref{eqn62} in \eqref{eqn19}, we obtain the directivity pattern as 
\begin{align}
    D(\hat{r}) &= 4\pi \frac{U(\hat{r})}{P_{rad}} = \frac{1}{P_{rad}}\frac{|I_0|^2}{2\pi}Z_f \left(\frac{\cos(\frac{\pi}{2}\cos\theta)}{\sin\theta}\right)^2\nonumber\\
    &= \frac{1}{\frac{|I_0|^2}{4\pi}Z_f(1.22)}\frac{|I_0|^2}{2\pi}Z_f \left(\frac{\cos(\frac{\pi}{2}\cos\theta)}{\sin\theta}\right)^2 = \frac{2}{1.22} \left(\frac{\cos(\frac{\pi}{2}\cos\theta)}{\sin\theta}\right)^2\nonumber\\
    &= 1.64 \left(\frac{\cos(\frac{\pi}{2}\cos\theta)}{\sin\theta}\right)^2. \label{eqn63}
\end{align}
The maximum value of the directivity pattern occurs at $\theta = \pi/2$ (see Fig. \ref{fig8})
\begin{equation}
D_0 = \left.D(\hat{r})\right|_{\theta = \pi/2} = 1.64 \left(\frac{\cos(\frac{\pi}{2}\cos(\pi/2))}{\sin(\pi/2))}\right)^2 = 1.64 \equiv 2.15 dB. \label{eqn64}
\end{equation}

\subsection{Radiation Resistance}\label{subsec4.4}
For an efficient antenna for which $R_L = 0$ (no power absorbed is lost as heat), the complex impedance is given by 
\begin{equation}
    Z_A = R_{rad} + j\omega L_A -j\frac{1}{\omega C_A}. \label{eqn65}
\end{equation}
With current $I_0$ flowing through antenna and voltage $V_A = I_0 Z_A$, the power absorbed from the source is all radiated ($P_L = 0$, see Eqn. \eqref{eqn27})
\begin{equation}
P_{rad} = \frac{1}{2}V_AI_0^* = \frac{1}{2}|I_0|^2Z_A = \frac{1}{2} |I_0|^2R_{rad} + j\frac{1}{2} |I_0|^2\omega L_A -j\frac{1}{2} |I_0|^2\frac{1}{\omega C_A}. \label{eqn66}
\end{equation}
Out of this so called `total radiated power', the `real radiated power' is $\frac{1}{2} |I_0|^2R_{rad}$. The remaining  `imaginary power' $\left(j\frac{1}{2} |I_0|^2\omega L_A -j\frac{1}{2} |I_0|^2\frac{1}{\omega C_A}\right)$
describes the flow of stored energy due to antenna capacitance ($C_A$) and inductance ($L_A$). For a half wave dipole, the capacitive energy and inductive energy stored at resonance are nearly equal and therefore (from Eqn. \eqref{eqn62}),
\begin{equation}
P_{rad} = \frac{1}{2} |I_0|^2R_{rad} = \frac{|I_0|^2}{4\pi}Z_f(1.22)~~~\Longrightarrow~~~R_{rad} = \frac{Z_f}{2\pi}(1.22) \simeq 73 \Omega.
\end{equation}
Half-wave dipoles are one of the most popular antennas because their radiation resistance is $73\Omega$, and therefore easy to connect to transmission lines of characteristic impedance 50, 75 or 100 $\Omega$, with little reflection due to impedance mismatch. In practice, the fundamental resonance of half wave dipoles does not occur exactly at $2l = 0.5\lambda$ but rather can happen anywhere from $0.3\lambda-0.48 \lambda$ depending upon their thickness and structure. For wire antennas shorter than half wavelength, the reactance is dominantly capacitive (negative) and for antennas longer than half wavelength, the reactance is dominantly inductive (positive). 

\section{Antenna Measurements}\label{sec5}
EM analysis of antenna, as discussed in sections \ref{sec2}, \ref{sec3} and \ref{sec4} are mathematically rigorous, time consuming, and prone to error. We therefore, mostly to manufacture more realistic antennas, rely on simulations. However, simulations also need to be verified. Therefore, antennas used in radio neutrino and cosmic ray experiments are characterized in a radio-quiet environment: an EM anechoic chamber. The instrument used to spec an antenna at an anechoic chamber is called a vector network analyzer (VNA). In this section, along with discussing different antenna characteristics, we show the measurements of an ARA \cite{1} bottom Vertically Polarized (Bottom VPol or BVPol) antenna. 

\subsection{Scattering (S) Parameters}\label{subsec5.1}
Assume that incident current and voltage waves ($V^+$ and $I^+$) from a Tx are reflected ($V^-$ and $I^-$) from a junction due to impedance mismatch while propagating through a transmission line with a characteristic impedance ($Z_0$) defined as 
\begin{equation}
\frac{V^+}{I^+} = Z_0 = \frac{V^-}{I^-}. \label{eqn68}
\end{equation}
S-parameters estimate the ratio of the normalized wave incident
\begin{equation}
a = \frac{V^+}{\sqrt{Z_0}} = \sqrt{Z_0}I^+
 \label{eqn69}
\end{equation}
on port `n' to the normalized wave emerging (i.e., reflected or scattered)
\begin{equation}
b = \frac{V^-}{\sqrt{Z_0}} = \sqrt{Z_0}I^-
\label{eqn70}
\end{equation}
from port `m' as
\begin{equation}
S_{mn} = \frac{\textcolor{black}{b_m}}{\textcolor{black}{a_n}} =  \frac{\textcolor{black}{V^-_m}}{\textcolor{black}{V^+_n}}\frac{\textcolor{black}{\sqrt{Z_{0n}}}}{\textcolor{black}{\sqrt{Z_{0m}}}} \label{eqn71}
\end{equation}
when all the ports $m \neq n$ are terminated with matched loads. For a complete characterization (gains and losses) of an antenna, a VNA with at least 2 ports is required. An $N$-port VNA can measure $N^2$ number of scattering parameters. For a 2-port VNA and a common transmission line, 
\begin{equation}
    {\bf \textcolor{black}{b} = \mathcal{S} \textcolor{black}{a}}~~\Longrightarrow~~\begin{pmatrix} b_1 \\ b_2 \end{pmatrix} = \begin{pmatrix}S_{11} & S_{12}\\ S_{21} & S_{22}\end{pmatrix} \begin{pmatrix} a_1 \\ a_2\end{pmatrix} \label{eqn72}
\end{equation}
such that,\\\\
\begin{equation}
    \mbox{Input~reflection~coefficient:}~~ \displaystyle{S_{11} = \frac{V_1^-}{V_1^+} = \left. \frac{b_1}{a_1} \right|_{a_2 = 0} } \label{eqn73}
\end{equation}
\begin{equation}
    \mbox{Forward ~transmission ~gain:}~~\displaystyle{S_{21} = \frac{V_2^-}{V_1^+} = \left. \frac{b_2}{a_1} \right|_{a_2 = 0}}\label{eqn74}
\end{equation}
\begin{equation}
    \mbox{Reverse ~transmission~ gain:}~~\displaystyle{S_{12} = \frac{V_1^-}{V_2^+}  = \left. \frac{b_1}{a_2} \right|_{a_1 = 0}} \label{eqn75}
\end{equation}
\begin{equation}
    \mbox{Output ~reflection ~coefficient:} ~~\displaystyle{S_{22} = \frac{V_2^-}{V_2^+} = \left. \frac{b_2}{a_2} \right|_{a_1 = 0}} \label{eqn76}
\end{equation}
If there is no active element (amplifier, etc.) in the measurement setup, then $S12 \equiv S21$. This is called `reciprocity' which is an inherent property of any linear, passive, time-invariant
EM device. A VNA can record complex S-parameters ($S = \Re(S) + j\Im(S)$), as $\Re$ and $\Im$ values or magnitude and phases. Typical contemporary VNAs employ a touchstone data format, saving data in `.sNp' files, with N being the number of VNA ports used. 

\subsubsection{Reflection Coefficient}\label{subsubsec5.1.1}
For an antenna connected with port 1 of a VNA, the reflection coefficient $\Gamma = S11$. If it is connected with port 2, then $\Gamma = S22$. In general,
\begin{equation}
    \Gamma = S ~~\Longrightarrow~~|\Gamma| = |S| = \sqrt{ \Re(S)^2+\Im(S)^2} \label{eqn77}
\end{equation}
\begin{figure}[h]
\centering
\includegraphics[width=0.75\textwidth]{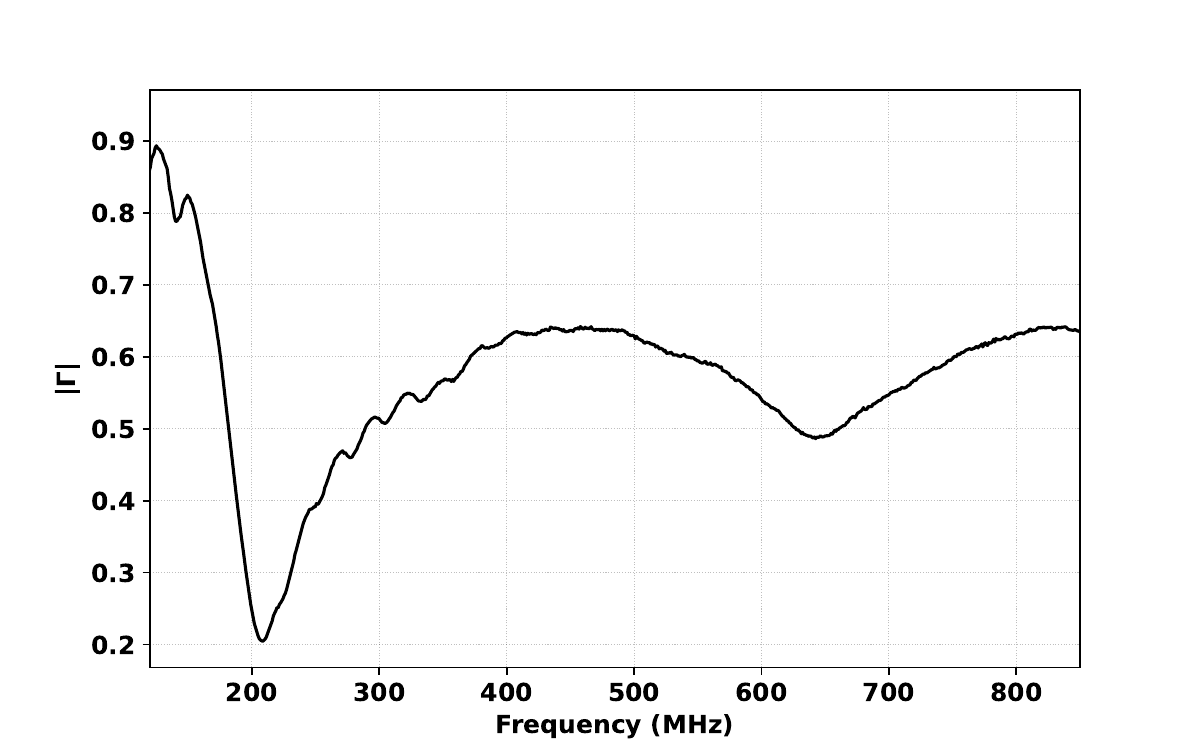}
\caption{Reflection coefficient of an ARA dipole antenna (Bottom VPol).}\label{fig1}
\end{figure}
Notice that, in Fig. \ref{fig1}, the reflection coefficient is minimal at $208$ MHz which corresponds to the fundamental frequency at which resonance occurs. The second minimum (first harmonic) occurs near $640$ MHz. Beyond this frequency, the dipole exhibits some quadrupole behavior (four lobes in zenith gain pattern as opposed to two lobes for a dipole; see Fig. \ref{fig8}).
\subsubsection{Antenna Impedance}\label{subsubsec5.1.2}
The antenna impedance can be extracted from S11 or S22 as 
\begin{equation}
    Z_A = Z_0 \left(\frac{1+S}{1-S}\right)\label{eqn78}
\end{equation}

\begin{figure}[h]
\centering
\includegraphics[width=0.75\textwidth]{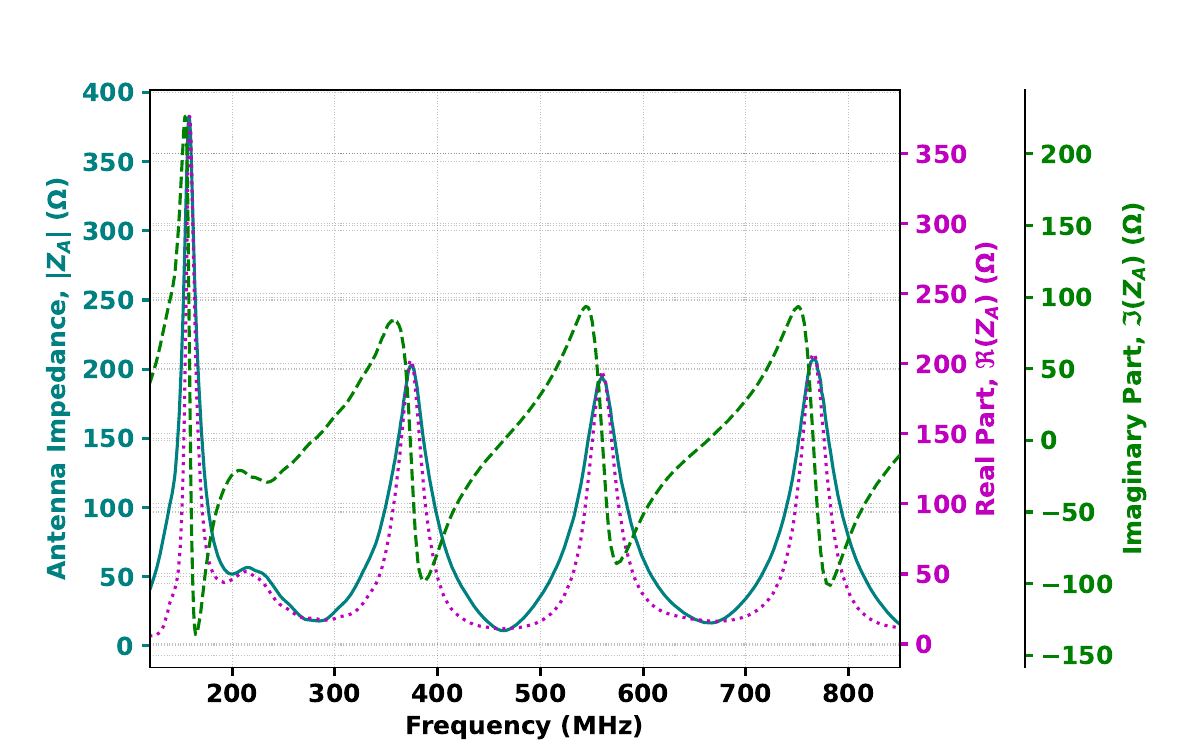}
\caption{Complex impedance of an ARA dipole antenna (Bottom VPol). Solid line is $|Z_A|$, dotted line is $\Re(Z_A)$, and dashed line is $\Im(Z_A)$.}\label{fig2}
\end{figure}

\subsubsection{Voltage Standing Wave Ratio (VSWR)}\label{subsubsec5.1.3}
This quantity, like the reflection coefficient, characterizes how well an antenna's impedance matches to the connecting transmission line, and can be calculated from S11 or S22. The total voltage through a transmission line is given by (from Eqns. \eqref{eqn73} and \eqref{eqn76})
\begin{equation}
    V = V^+ + V^- = V^+\left(1 + \frac{V^-}{V^+} \right) = V^+\left(1 + \Gamma \right).\label{eqn79}
\end{equation}
VSWR is defined as the ratio of maximum to minimum of the total voltage magnitude,
\begin{equation}
    \mathrm{VSWR} = \frac{|V|_{max}}{|V|_{min}} = \frac{1+|\Gamma|}{1-|\Gamma|} = \frac{1+|S|}{1-|S|}. \label{eqn80}
\end{equation}
Notice that VSWR can have a minimum value of 1 corresponding to no reflection and a maximum of $\infty$ corresponding to complete reflection.
\begin{figure}[h]
\centering
\includegraphics[width=0.75\textwidth]{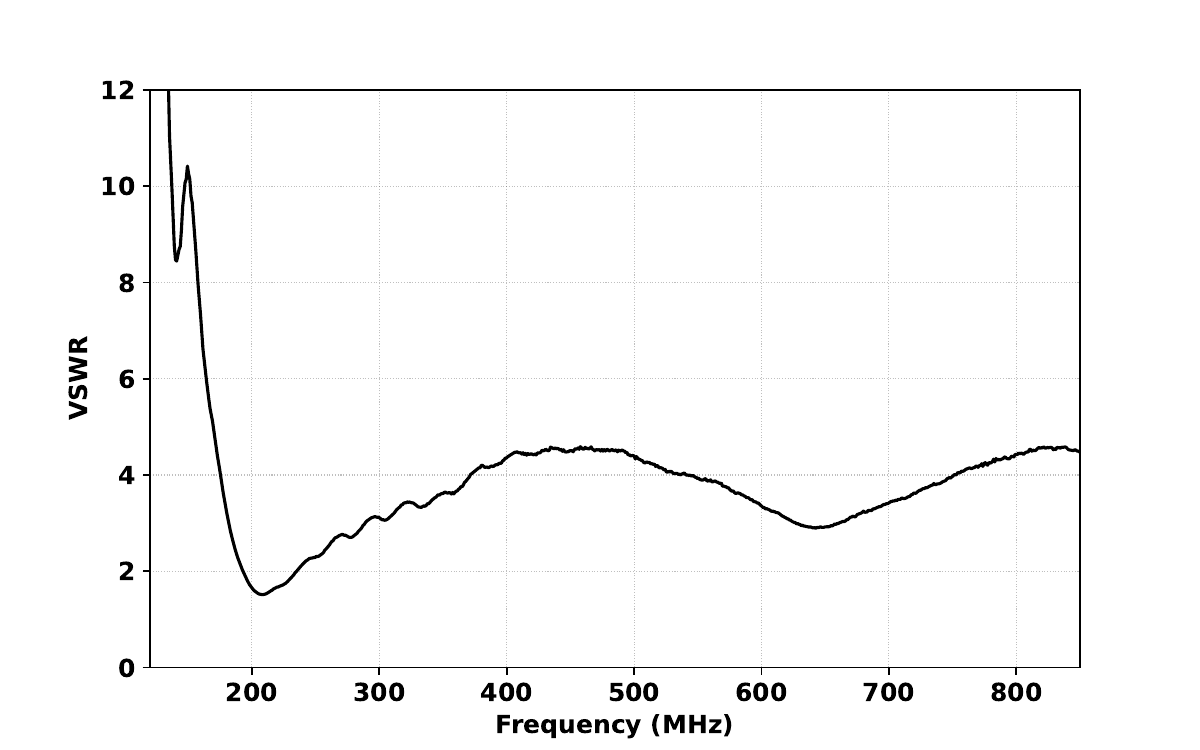}
\caption{Voltage Standing Wave Ratio (VSWR) of an ARA dipole antenna (Bottom VPol).}\label{fig3}
\end{figure}

\subsubsection{Return Loss (RL)}\label{subsubsec5.1.4}
Return loss is a measure of the ratio of reflected power to the incident power. In practice, we express it in dB as
\begin{equation}
    RL = - 10\log_{10}{\left(\frac{P_{ref}}{P_{in}}\right)} = -20\log_{10}{|\Gamma|}. \label{eqn81}
\end{equation}
Note that the negative sign indicates loss. For example, a return loss of 10 dB corresponds to {10\%} of the power reflected and 20 dB return loss corresponds to 1\% power reflection. The larger the value of return loss in dB, the smaller the impedance mismatch and therefore better the antenna. 
\begin{figure}[h]
\centering
\includegraphics[width=0.75\textwidth]{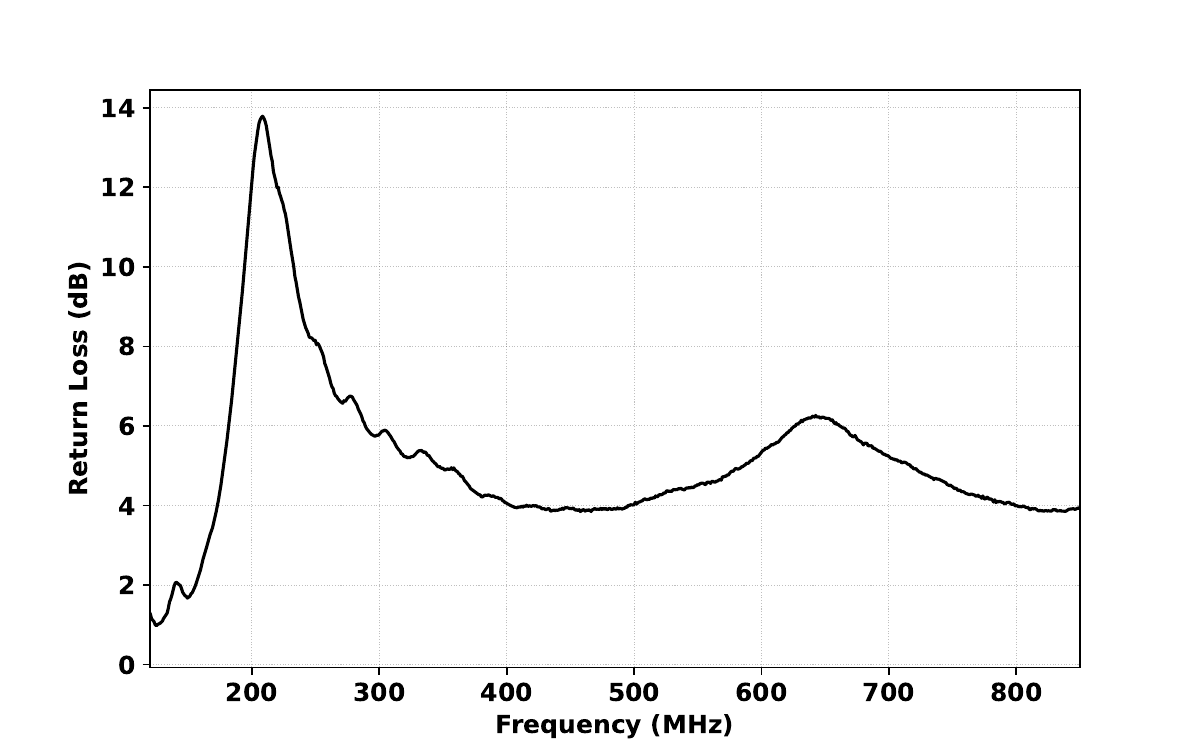}
\caption{Return Loss (in dB) of an ARA dipole antenna (Bottom VPol).}\label{fig4}
\end{figure}
Here, the ARA BVPol antennas have minimum loss at resonance frequency, $208$ MHz as shown in Figs. \ref{fig1}, \ref{fig3} and \ref{fig4}.
\subsubsection{Group Delay}\label{subsubsec5.1.5}
In radio astroparticle physics experiments, we work with pulsed signals  with a large bandwidth, e.g., $50-1000$ MHz. If the group of frequencies travel at different phase velocities, delays are evident in the time-domain pulses. This group delay mostly depends on the inductive and capacitive properties of an antenna as a function of frequency, and is defined as: 
\begin{equation}
    \tau_g = -\frac{\partial \Phi}{\partial \omega} = -\frac{1}{2\pi}\frac{\partial \Phi}{\partial f}; ~~~\Phi=\tan^{-1}\left(\frac{\Im(S)}{\Re(S)}\right) \label{eqn82}
\end{equation}
\begin{figure}[h]
\centering
\includegraphics[width=0.75\textwidth]{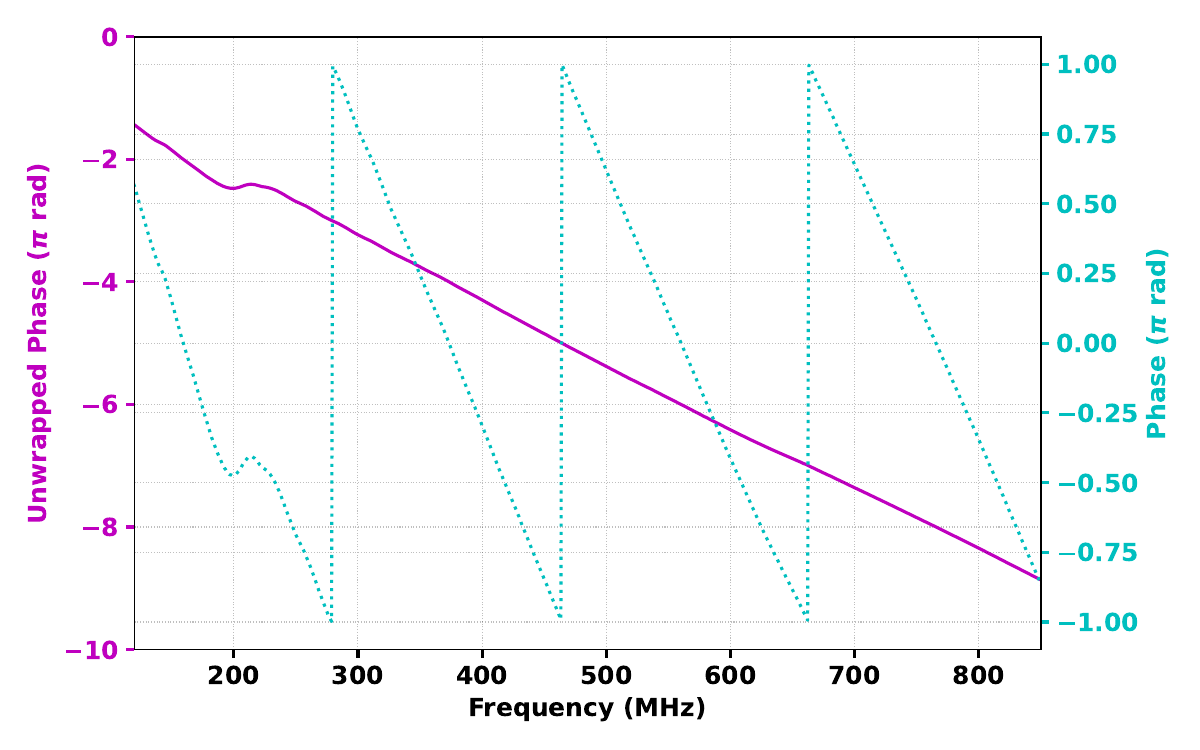}
\caption{Wrapped (dashed line) and unwrapped (solid line) phases of an ARA dipole antenna (Bottom VPol).}\label{fig5}
\end{figure}
The phases of antennas are wrapped (periodic) with frequencies spanning from $-\pi$ rad to $+\pi$ rad as shown in Fig. \ref{fig5}. To calculate the group delay in Fig. \ref{fig6}, the phase first needs to be unwrapped. The negative sign in Eqn. \eqref{eqn82} indicates that the unwrapped phase decreases with frequency. Notice that the unwrapped phase is almost linear throughout the bandwidth which results in a constant group delay except for at resonance frequency (Fig. \ref{fig6}) where the phase seems to have a point of inflection (Fig. \ref{fig5}). 
\begin{figure}[h]
\centering
\includegraphics[width=0.75\textwidth]{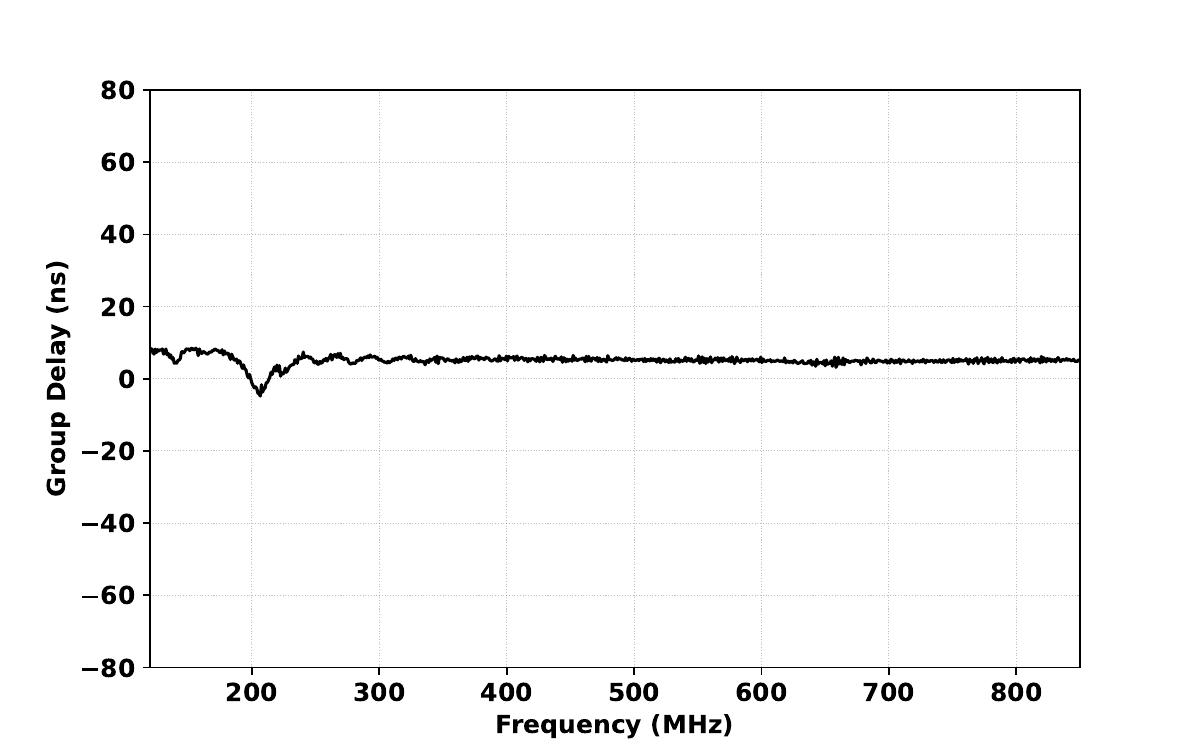}
\caption{Group delay (in ns) of an ARA dipole antenna (Bottom VPol).}\label{fig6}
\end{figure}

\subsubsection{Impulse Response}\label{subsubsec5.1.6}
Every system has a response to an input impulse (e.g., a scaled delta function). In the time-domain, the output ($y(t)$) is a convolution of the input ($x(t)$) and the impulse response ($h(t)$) of that system. The transfer function ($H(\omega)$) is related to impulse response in frequency space (through Fourier transform, $\mathcal{F}$):
\begin{equation}
    y(t) = h(t)*x(t) = \int_{-\infty}^{\infty} h(\tau) x(t - \tau) d\tau; ~~~H(\omega) = \int_{-\infty}^{\infty} h(t) e^{-j\omega t} dt
\end{equation}
S-parameters are somewhat similar to the transfer functions of antennas. Therefore,
\begin{align}
    S(\omega) = \int_{-\infty}^{\infty} h(t) e^{-j\omega t} dt ~~\Longleftrightarrow~~h(t) = \frac{1}{2\pi} \int_{-\infty}^{\infty} S(\omega) e^{+j\omega t} d\omega  = \mathcal{F}^{-1}(S)
\end{align} 
\begin{figure}[h]
\centering
\includegraphics[width=0.75\textwidth]{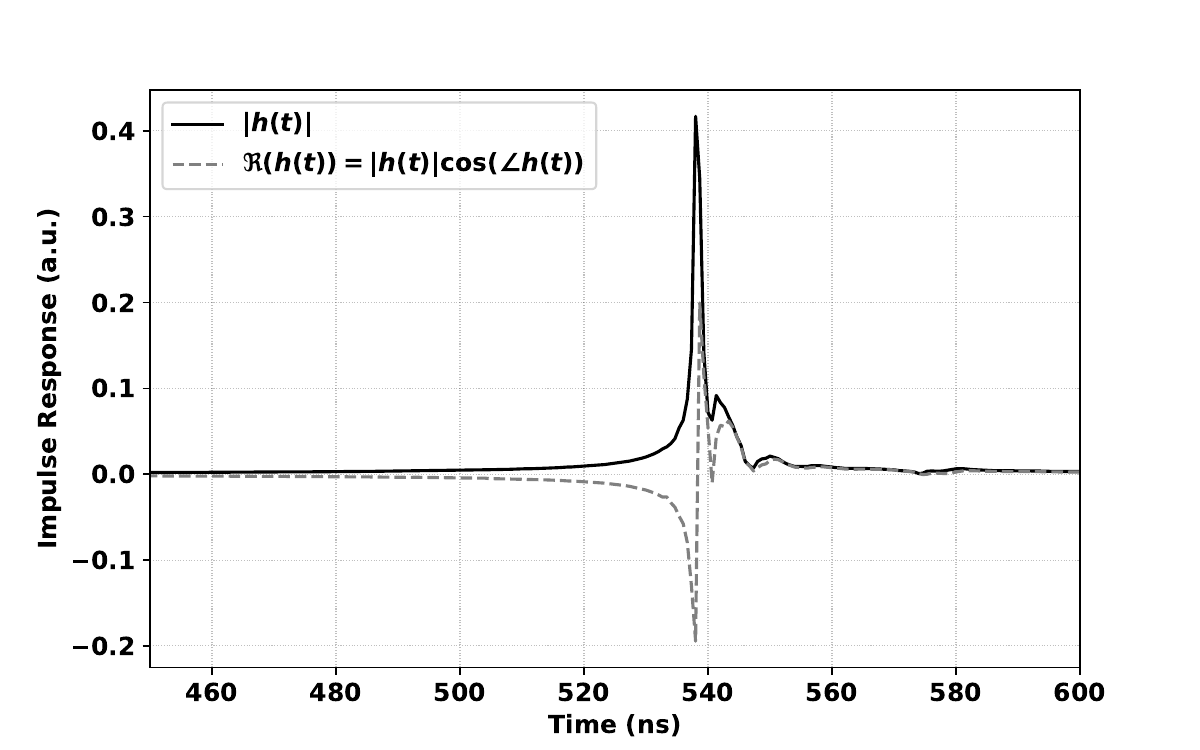}
\caption{Impulse Response of an ARA dipole antenna (Bottom VPol).}\label{fig7}
\end{figure}
In general, the impulse response is sharp ($<10$ ns) for dipole antennas, making them suitable for the detection of sharp Askaryan radiation pulses \cite{1}\cite{2}. In case of Log Periodic Dipole Array (LPDA) antennas, this width can be as large as $>50$ ns.

\subsection{Gain Calculation using the Friis Transmission Equation}\label{subsec5.2}
\begin{figure}[h]
\centering
\includegraphics[width=0.5\textwidth]{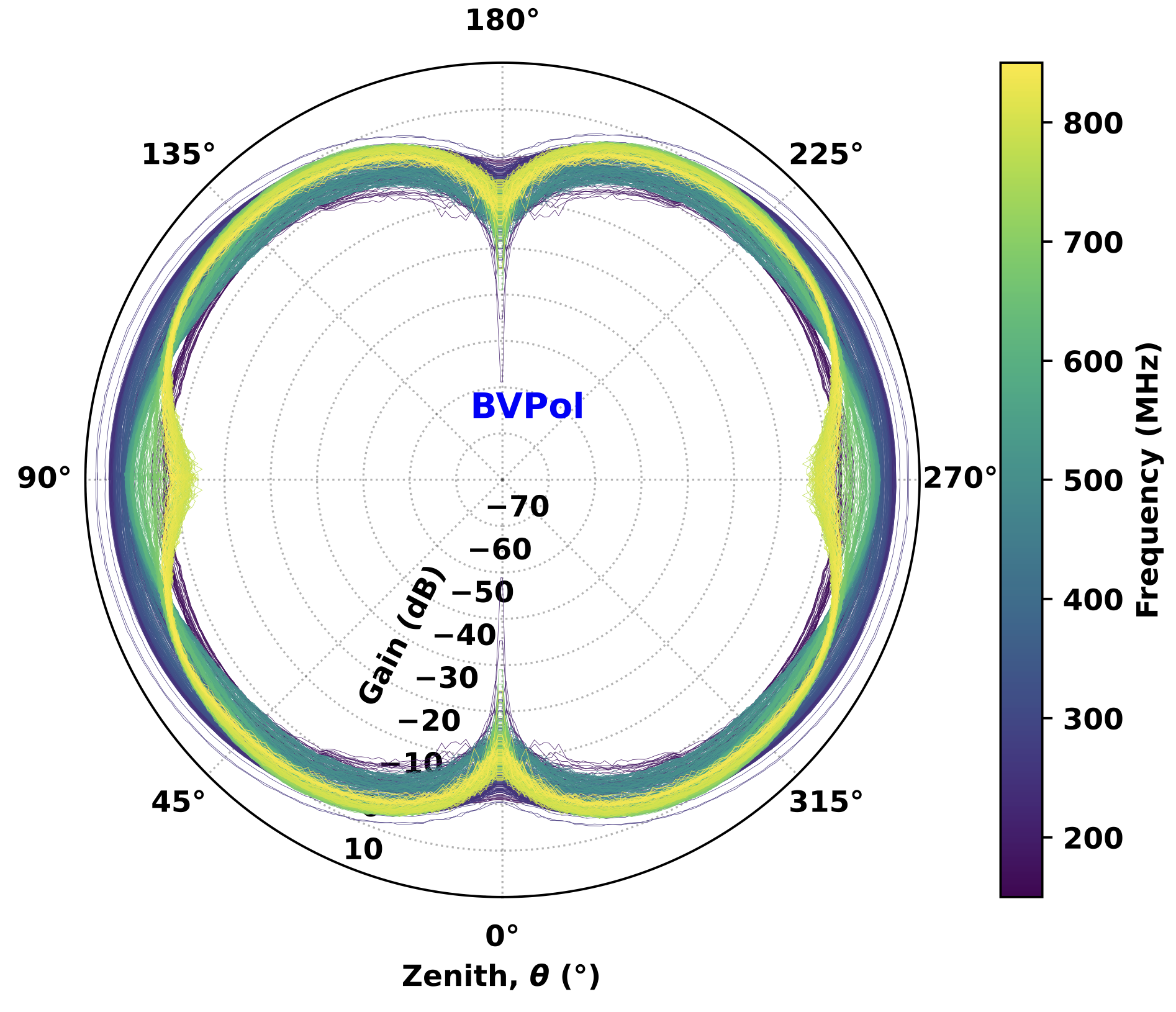}~\includegraphics[width=0.5\textwidth]{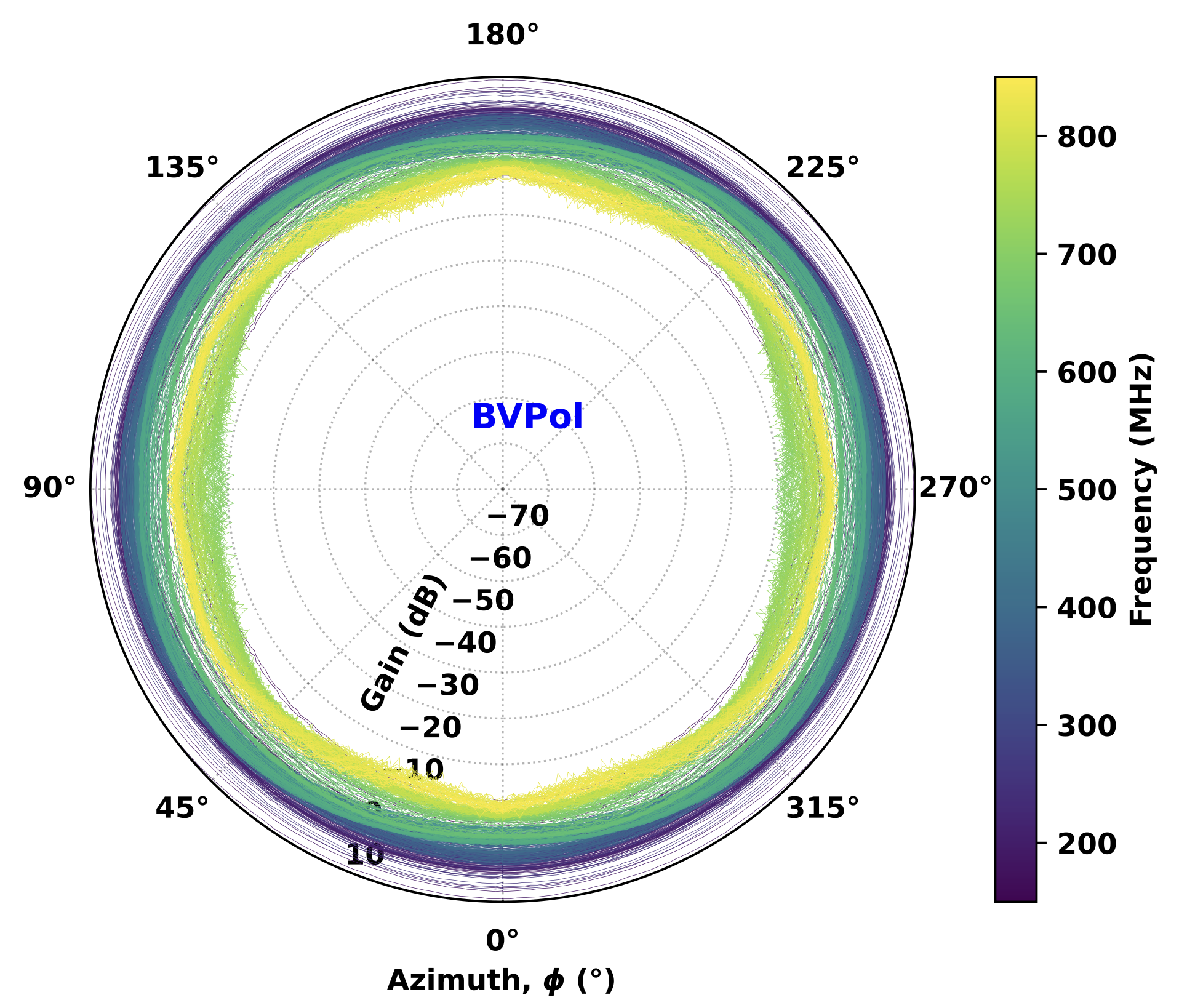}
\caption{Realized gain patterns of an ARA dipole antenna (Bottom VPol) in zenith (left) and azimuth (right) planes.}\label{fig8}
\end{figure}

For a stationary Tx (reference antenna with realized gain pattern, $G^\theta_{ref}$) with available power, $P_t$ and a rotating Rx (`antenna under test' (aut) with realized gain pattern, $G^\theta_{aut}$) absorbing $P_r$ power, both set in co-polarization configuration ($PLF = 1$), the Friis transmission equation (from Eqn. \eqref{eqn48}) is

\begin{equation}
    \frac{P_r}{P_t} = \left(\frac{\lambda}{4\pi R}\right)^2 G^\theta_{ref} G^\theta_{aut}. \label{eqn85}
\end{equation}
For identical Tx and Rx at $\theta = 0^\circ$, 
\begin{align}
    G^{\theta=0}_{ref} = G^{\theta=0}_{aut},~~~\&~~~
    \frac{P_r}{P_t} = \left(\frac{V_r}{V_t}\right)^2 \simeq |S_{21}|^2. \label{eqn86}
\end{align}
\begin{equation}
    \therefore ~~~|S_{21}|^2 = \left(\frac{\lambda}{4\pi R}\right)^2 G^{\theta=0}_{ref} G^{\theta=0}_{aut}~~\Longrightarrow ~~ G^{\theta=0}_{ref} = \left(\frac{4\pi R}{\lambda}\right)|S_{21}|. \label{eqn87}
\end{equation}
For identical Tx and Rx at different angles, we extract the realized gain of the `aut' as
\begin{equation}
    G^{\theta}_{aut} = \frac{1}{G^{\theta=0}_{ref}}\left(\frac{4\pi R}{\lambda}\right)^2|S_{21}|^2. \label{eqn88}
\end{equation}
If the Tx and Rx are not identical, this method would not be ideal for gain extraction from S21. But if the gain pattern of an antenna is fully known, it can be used as a reference antenna to spec the gain pattern of any other antenna. For example, we can use two identical LPDA antennas to scan the gain pattern of one of the LPDAs. Then we use that LPDA as a reference antenna to obtain the gain pattern of an ARA Bottom VPol antenna as shown in Fig. \ref{fig8}.
Often for a dipole antenna, we also denote the plane swept in zenith as the E-plane (as it contains the electric field vector; see Eqn. \eqref{eqn53}) and the plane swept by azimuth angle as H-plane (as it contains the magnetizing field vector ; see Eqn. \eqref{eqn54}). More details and demonstrations of E \& H planes for dipole and spiral antennas can be found in \cite{6}. For ARA BVPol antennas, at higher frequencies (mentioned in section \ref{subsubsec5.1.1}), some quadrupole behavior also appears, as can be seen in the zenith pattern of in Fig. \ref{fig8}. The azimuth pattern (right plot in Fig. \ref{fig8}) should be ideally symmetric. Some degree of asymmetry arises at higher frequencies mostly due to through cables in the BVPol antennas and measurement uncertainties.

\subsection{Effective Lengths}\label{subsec5.3}
Most radio neutrino and cosmic ray detectors record voltage vs time traces ($V(t)$) in their data acquisition systems. However, EM waves propagate as fields. When a time varying field ($\Vec{E}(t)$) is intercepted by a radio antenna, it is convolved with the vector effective height ($\Vec{H}(t)$) of the antenna to produce the observed scalar voltage \cite{7}. Writing the antenna response in the frequency domain: 
\begin{equation}
    V(t) = \Vec{H}(t) *\Vec{E}(t) ~~~\xrightleftharpoons[\mathcal{F}^{-1}]{\mathcal{F}} ~~~ \mathcal{V}(\omega) = \Vec{\mathcal{H}}(\omega)\cdot \Vec{\mathcal{E}}(\omega). \label{eqn89}
\end{equation}
Absolute value of the components ($i = \theta, \phi$) of the vector effective height \cite{5}\cite{7}\cite{8} can be expressed in terms of the antenna gain as 
\begin{equation}
    |\Vec{\mathcal{H}}_i (\omega)| = \lambda \sqrt{\frac{\Re(Z_A)}{\pi Z_f} G_i(\omega)}. \label{eqn90}
\end{equation}
Corresponding to realized gain, calculated from S21 in section \ref{subsec5.2}, the realized vector effective height \cite{7}\cite{8} can be expressed as 
\begin{equation}
    |\Vec{\mathcal{H}}^r_i (\omega)| = \lambda \sqrt{\frac{Z_0}{4\pi Z_f} G^r_i(\omega)} = \lambda \sqrt{\frac{Z_0}{4\pi Z_f} G_i(\omega) (1 - |\Gamma_A (\omega)|^2)}. \label{eqn91}
\end{equation}
\begin{figure}[h]
\centering
\includegraphics[width=0.75\textwidth]{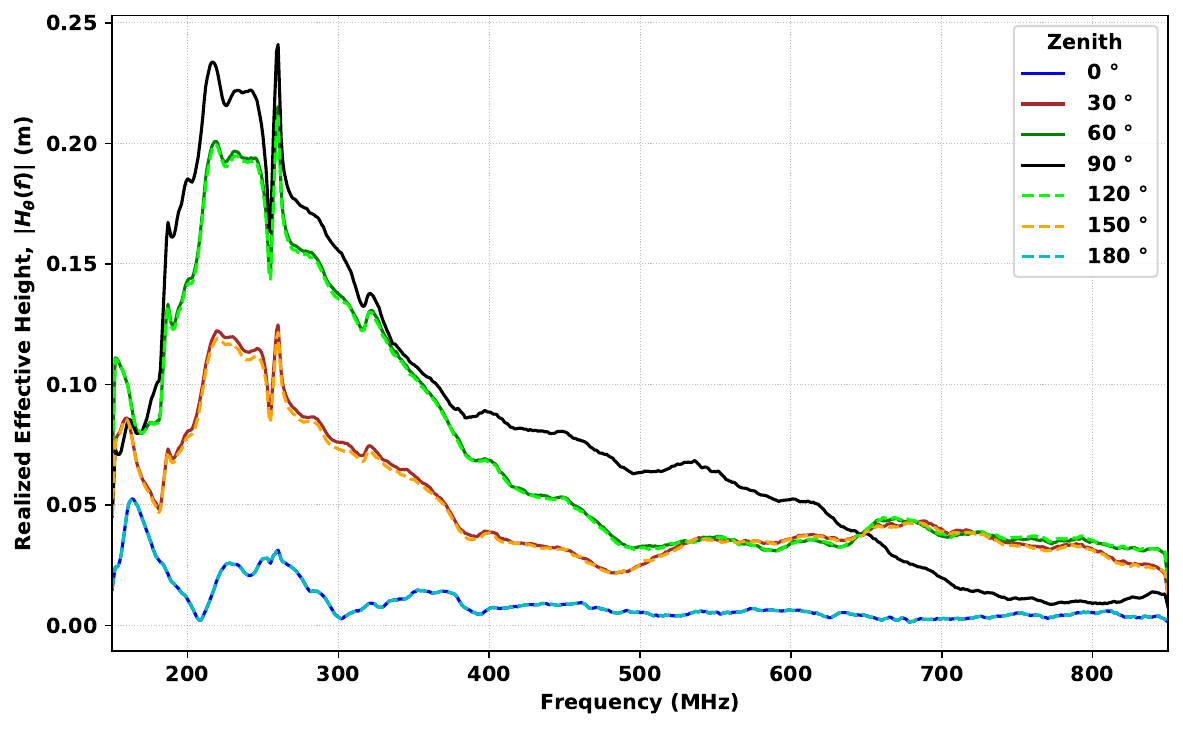}\\
\includegraphics[width=0.75\textwidth]{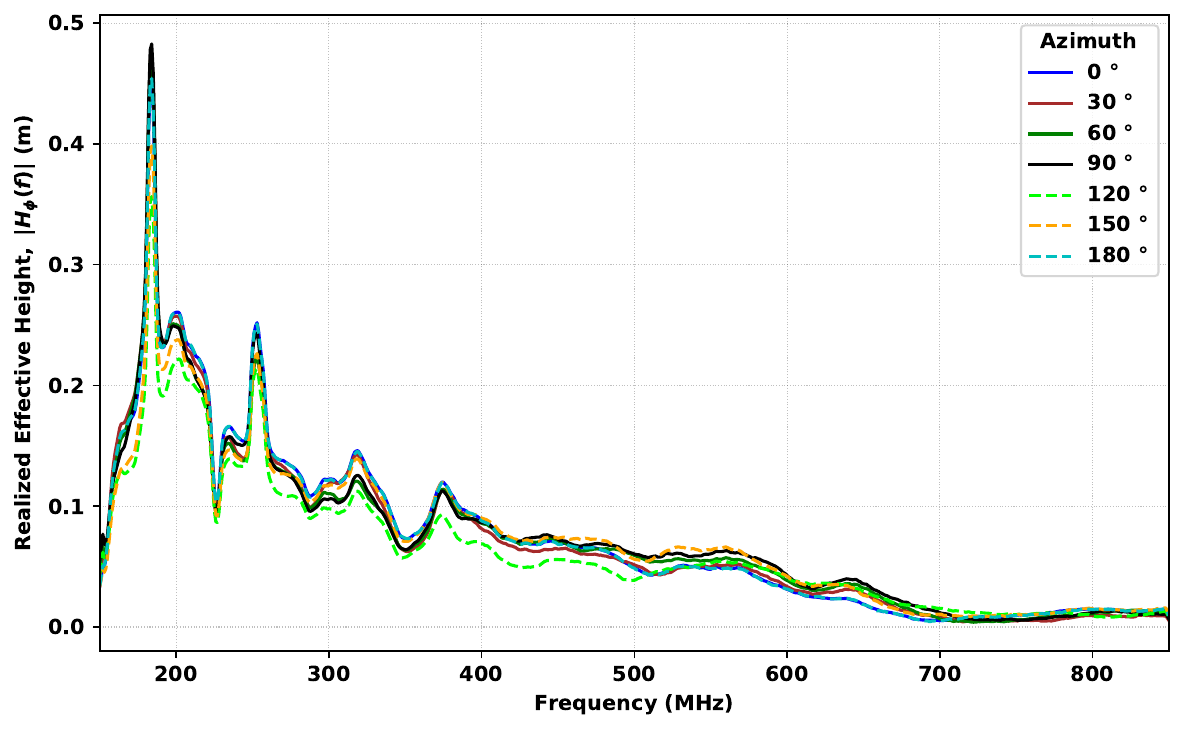}
\caption{Realized effective height of an ARA dipole antenna (Bottom VPol) in zenith (top) and azimuth (bottom) planes.}\label{fig9}
\end{figure}Notice that the azimuthally realized effective height, shown in Fig. \ref{fig9}, is symmetric over $\phi$ due to the azimuthally symmetric gain pattern, shown in Fig. \ref{fig8}. 

\section{Summary and Conclusions}\label{sec6}
We briefly discussed the underlying physics of antennas both in general and for radio antennas in particular, highlighting the case of a half-wave dipole. We defined the important antenna characteristics in terms of S-parameters.  We also presented measurements of the ARA Bottom VPol antenna at the University of Kansas anechoic chamber.\\\\ 
Radio antennas are fascinating, and understanding their characteristics is vital for the UHEN and UHECR experiments that use them. The physics and mathematics behind them are rich as well as subtle. Measurements in an anechoic chamber with a vector network analyzer facilitate understanding of radio antennas. However, antenna characteristics change significantly in a medium with depth dependent refractive index, e.g., polar ice. Therefore, precise modeling of antenna for radio neutrino and cosmic ray experiments requires sophisticated simulation tools. 

\newpage
\section*{Acknowledgments}\label{sec7}
The author would like to thank Professor Jim Stiles, EECS Department, University of Kansas who teaches the course EECS 721 (Antennas); many concepts are inspired from his lectures. The author is also very grateful to NSF's IEI grant 2019597, ARA grant 2310096, RNO-G grant 2310126, and RET grant 2012989 for their financial support.

\bibliographystyle{unsrt}

\end{document}